\documentclass[%
 reprint,
superscriptaddress,
nofootinbib,
 amsmath,amssymb,
 aps,
]{revtex4-1}
\usepackage{graphics,epsfig,psfrag}
\usepackage{color}
\usepackage{graphicx}
\usepackage{dcolumn}
\usepackage{bm}

\begin{document}

\preprint{APS/123-QED}

\title{Influence of a plasma on the shadow of a spherically 
      symmetric black hole}

\author{Volker Perlick}
\email{perlick@zarm.uni-bremen.de}
\affiliation{ZARM, University of Bremen, 28359 Bremen,
Germany}

\author{Oleg Yu. Tsupko}
\email{tsupko@iki.rssi.ru}
\affiliation{Space Research Institute of Russian Academy of Sciences, Profsoyuznaya 84/32, Moscow 117997, Russia}

\author{Gennady S. Bisnovatyi-Kogan}
\email{gkogan@iki.rssi.ru}
\affiliation{Space Research Institute of Russian Academy of Sciences, Profsoyuznaya 84/32, Moscow 117997, Russia}

\date{\today}

\begin{abstract}
We analytically calculate the influence of a plasma on the 
shadow of a black hole (or of another compact object). 
We restrict to spherically symmetric and static situations, where
the shadow is circular. The plasma is assumed to be non-magnetized
and pressure-less. We derive the general formulas for a spherically 
symmetric plasma density on an unspecified spherically symmetric and 
static spacetime. Our main result is an analytical
formula for the angular size of the shadow. As a plasma is a dispersive
medium, the radius of the shadow depends on the photon frequency. The
effect of the plasma is significant only in the radio regime.
The formalism applies not only to black holes
but also, e.g., to wormholes. As examples for the underlying spacetime 
model, we consider the Schwarzschild spacetime and the Ellis wormhole. In 
particular, we treat the case that the plasma is in radial free fall 
from infinity onto a Schwarzschild black hole.
We find that for an observer far away from a Schwarzschild black
hole the plasma has a decreasing effect on the size of the shadow.
The perspectives of 
actually observing the influence of a plasma on the shadows of 
supermassive black holes are discussed.
\begin{description}
\item[PACS numbers] 04.20.-q -- 98.62.Sb -- 98.62.Mw -- 98.35.Jk


\end{description}
\pacs{?????? - ??????}
\end{abstract}

\pacs{?????? - ??????}
\maketitle


\section{Introduction}\label{sec:intro}

In the last twenty years increasing evidence has been found for the existence
of a supermassive black hole at the center of our galaxy. It is expected that a
distant observer should ``see'' this black hole as a dark disk in the sky which is
known as the ``shadow''. It is sometimes said that the shadow
is an image of the event horizon, and the size of the shadow is
estimated by the angle under which the observer would see the 
horizon according to Euclidean geometry. Actually, the 
boundary of the shadow corresponds to light rays that asymptotically 
approach the photon sphere (at $r=3M$ in the Schwarzschild case) and 
not the horizon (at $r=2M$ in the Schwarzschild case). Moreover, 
light rays do not follow straight lines in Euclidean geometry 
but they are bent. For these two reasons, the angular diameter of
the shadow is actually bigger than the naive Euclidean estimate
suggests. For the black hole at the center of our galaxy, it
amounts to about 53 $\mu$as whereas the Euclidean estimate gives
only about 20 $\mu$as. At present, two projects are under way to  observe 
this shadow which would give important information on the compact object at 
the center of our galaxy. These projects, which are going to use 
(sub)millimeter VLBI observations with radio telescopes distributed over
the Earth, are the Event Horizon Telescope (http://eventhorizontelescope.org)
and the BlackHoleCam (http://blackholecam.org). 

On the theoretical side, the shadow is defined as the region of the observer's
sky that is left dark if there are light sources distributed everywhere but not
between the observer and the black hole. For a non-rotating black hole, the
shadow is a circular disk. For a Schwarzschild black hole the angular diameter 
of the shadow was calculated, as a function of the mass of the black hole and of the
radius coordinate where the observer is situated, by Synge \cite{Synge1966}.
(Synge did not use the word ``shadow''. He calculated what he called the
``escape cones'' of light. However, the complement of the opening angle of 
the escape cone at the observer gives exactly the diameter of the shadow.) 
For a rotating black hole the shadow is no longer circular. The first to correctly 
calculate the shape of the shadow of a Kerr black hole was 
Bardeen~\cite{Bardeen1973}. More generally, the shape and the size of the 
shadow can be calculated analytically for the whole class of Pleba{\'n}ski-Demia{\'n}ski 
space\-times~\cite{GrenzebachPerlickLaemmerzahl1914,GrenzebachPerlickLaemmerzahl1915}.
All these calculations are based on the assumption that light propagates along
lightlike geodesics, without being influenced directly by matter. If such an 
influence is to be taken into account, one usually resorts to numerical 
calculation. In particular, Falcke, Melia and Agol \cite{FalckeMeliaAgol2000}
have numerically simulated the visual appearance of the black hole at the 
center of our galaxy, assuming that it is a Kerr black hole, with scattering 
and the presence of emission regions between the observer and the black hole 
taken into account (at 0.6 and 1.3 mm wavelengths). 
Sophisticated ray tracing programs have been written for 
producing realistic images of a black hole surrounded by an accretion disk, 
e.g. for the movie \emph{Interstellar}. The numerical techniques used for 
this movie are described in detail, along with a review of earlier work, 
by James et al. \cite{JamesTunzelmannFranklinThorne2015}.

While the purely geometric calculation of the shadow can be done analytically,
at least for spacetimes in which the equation for lightlike geodesics is 
completely integrable, virtually all of the work that takes the influence
of matter into account was based on numerics so far. In this paper we
want to take the first steps towards an investigation of the shadow in
matter based on analytical calculations. It is clear that this can be done
only for highly idealized situations, but nonetheless we think that it is 
worthwile to do so. In contrast to numerical simulations, which can depict
the situation only for a particular choice of the parameters involved, 
analytical results demonstrate how exactly an effect depends on these
parameters and they bring out the general features more clearly. Also,
analytical results are useful for testing numerical codes. In this paper
we begin with the simplest non-trivial case: We want to analytically 
calculate the influence of a non-magnetized pressure-less
plasma on the size of the shadow of a non-rotating black hole.

A plasma is a dispersive medium, i.e., the light rays deviate from lightlike 
geodesics in a way that depends on the frequency. The Hamiltonian for
the light rays can be derived from Maxwell's equations where the source 
of the electromagnetic field consists of two charged fluids, one modelling 
the ions and the other the electrons. For a plasma on a curved background, 
the transition from Maxwell's equations to ray optics has to be done by 
a two-scale method. A rigorous derivation of the Hamiltonian for light rays 
was provided by Breuer and Ehlers \cite{BreuerEhlers1980,BreuerEhlers1981} 
who considered a magnetized pressure-less plasma. 
For the much simpler case of a non-magnetized pressure-less plasma, 
a similar derivation can be found in Perlick \cite{Perlick2000b}.  In the
latter case the equation of light rays can be fully characterized by
a scalar, i.e. direction-independent, index of refraction which is 
a function of the spacetime point and of the frequency. Then the resulting 
equation of motion for light rays belongs to a class that was studied
in the text-book by Synge \cite{Synge1960}.  The resulting formula for 
the deflection  angle of light in a plasma whose density is a function
of the radius coordinate was calculated in the Schwarzschild spacetime
(and more generally in the equatorial plane of the Kerr metric) by Perlick
\cite{Perlick2000b}. The same result was found and further discussed by
Bisnovatyi-Kogan and 
Tsupko \cite{BKTs2009}, \cite{BKTs2010}, \cite{TsBK2013}.
Morozova et al. \cite{MorozovaAhmedovTursunov2013}
generalized the calculation to the Kerr metric off the
equatorial plane, assuming that the black hole is slowly rotating.
For recent suggestions of how the effects of a plasma on light rays could
actually be observed we refer to Er and Mao \cite{ErMao2013} and
to Rogers \cite{Rogers2015}. 

In this paper we calculate the angular diameter of the shadow for a 
non-rotating black hole that is surrounded by a non-magnetized 
pressure-less plasma. Although we are mainly interested 
in the Schwarzschild case, in large parts of the paper we work in an 
unspecified spherically symmetric and static spacetime, so the results
can be applied, e.g., also to charged black holes or to wormholes.

In our work the plasma manifests itself as a refractive and dispersive medium, 
which leads to a change of the light rays in comparison with vacuum. As a result, 
the geometrical size of the shadow is changed. We do not take into account the 
processes of absorption and scattering of the photons by plasma electrons. Also,
we neglect the gravitational field of the plasma particles.

The paper is organized as follows. In Section \ref{sec:eom} we work out the 
equation of motion for light rays in a plasma on a spherically symmetric and 
static spacetime. In Section \ref{sec:circular} we determine the circular 
light orbits which are of crucial relevance for the formation of the shadow.
In Section \ref{sec:shadow} we calculate the angular diameter of the shadow.
We specify the results in Section \ref{sec:thin} to the case of a 
low-density plasma and in Section \ref{sec:accretion} to the case that the
plasma has the particular density profile that corresponds to spherically 
symmetric accretion of a dust onto a Schwarzschild black hole. 
In the same section we also discuss the perspectives
of actually observing the influence of a plasma on the shadow of
a supermassive black hole.

We use units such that $G=c=1$, so the Schwarzschild radius is 
$R_S=2M$. Our choice of signature is $\{-,+,+,+\}$.
We use Einstein's summation convention for latin indices
which take the values $i,k,$ $\dots$ $= t, r, \vartheta, \varphi$.

\section{Equations of motion for light rays in a non-magnetized 
plasma}\label{sec:eom}

We consider a spherically symmetric and static metric
\begin{equation}\label{eq:g}
g_{ik}dx^{i} dx^{k} 
= - A(r) dt^2 + B(r) dr^2 +
D(r) \big( d \vartheta ^2 + \mathrm{sin} ^2 \vartheta \, 
d \varphi ^2 \big) \, ,
\end{equation}
where $A(r)$, $B(r)$ and $D(r)$ are positive.
We are mainly interested in the case of a Schwarzschild
black hole but for the time being there is
no need for specifying the metric any further.

We assume that the spacetime is filled with a non-magnetized
cold plasma whose electron plasma frequency $\omega _p$ is a function 
of the radius coordinate only, 
\begin{equation}
\omega_p(r)^2 = \frac{4\pi e^2}{m} N(r) \, .
\end{equation}
Here $e$ is the charge of the electron, $m$ is the electron mass, and 
$N(r)$ is the number density of the electrons in the plasma. 
The refraction index $n$ of this plasma depends on the radius coordinate
$r$ and on the frequency $\omega$ of the photon as it is measured
by a static observer,
\begin{equation}
n( r , \omega ) ^2 = 1 - \frac{\omega_p(r)^2}{\omega^2} \, .
\end{equation}
Because of the spherical symmetry we may restrict to the equatorial 
plane $\vartheta = \pi /2$, $p_{\vartheta}=0$. Then the Hamiltonian 
for light rays in the plasma reads 
\begin{equation}\label{eq:H}
H \, = \, \dfrac{1}{2} \left( g^{ik} p_{i} p_{k}
+ \omega _p (r) ^2 \right) \, = \, \end{equation}
\[
= \, \dfrac{1}{2}
\left( - \dfrac{p_t^2}{A(r)} + \dfrac{p_r^2}{B(r)}
+ \dfrac{p_{\varphi}^2}{D(r)}+
\omega _p (r) ^2  \right) \, .
\]
For a derivation of the Hamiltonian (\ref{eq:H}) from Maxwell's equations 
with a two-fluid source we refer to the literature given in the introduction.

The light rays are the solutions to Hamilton's equations
\begin{equation}\label{eq:Ham0}
\dot{p}{}_i = -\dfrac{\partial H}{\partial x^i}
\, , \quad
\dot{x}{}^i = \dfrac{\partial H}{\partial p_i}
\end{equation}
which in this case read
\begin{equation}\label{eq:Ham1}
\dot{p}{}_t \, = \, - \, \dfrac{\partial H}{\partial t} \, = \, 0 \, ,
\end{equation}
\begin{equation}\label{eq:Ham2}
\dot{p}{}_{\varphi} \, = \, 
- \, \dfrac{\partial H}{\partial \varphi} \, = \, 0  \, ,
\end{equation}
\begin{gather}\label{eq:Ham3}
\dot{p}{}_r \, = \, - \, \dfrac{\partial H}{\partial r} \, = \,
\\
\nonumber
\dfrac{1}{2} \, \Big( - \dfrac{p_t^2A'(r)}{A(r)^2}
+ \dfrac{p_r^2B'(r)}{B(r)^2} + \dfrac{p_{\varphi}^2D'(r)}{D(r)^2}
- \dfrac{d }{dr} \omega _p (r) ^2\Big)  \, ,
\end{gather}
\begin{equation}\label{eq:Ham4}
\dot{t} \, = \, \dfrac{\partial H}{\partial p_t} \, = \, 
- \, \dfrac{p_t}{A(r)}   \, ,
\end{equation}
\begin{equation}\label{eq:Ham5}
\dot{\varphi} \, = \, \dfrac{\partial H}{\partial p_{\varphi}} \, = \, 
\dfrac{p_{\varphi}}{D(r)}   \, ,
\end{equation}
\begin{equation}\label{eq:Ham6}
\dot{r} \, = \, \dfrac{\partial H}{\partial p_r} \, = \, 
\dfrac{p_r}{B(r)}   \, ,
\end{equation}
with $H=0$, i.e.
\begin{equation}\label{eq:H0}
0 \, = \, - \dfrac{p_t^2}{A(r)} + \dfrac{p_r^2}{B(r)}
+ \dfrac{p_{\varphi}^2}{D(r)}+
\omega _p (r) ^2  \, .
\end{equation}
Here a dot means differentiation with respect to an affine parameter 
$\lambda$, and a prime means differentiation with respect to $r$.

>From (\ref{eq:Ham1}) and (\ref{eq:Ham2}) it follows that $p_t$ and 
$p_\varphi$ are constants of motion. We write $\omega _0 := - p_t$. 
If $\omega _0$ has been fixed and if $A(r) \to 1$ for $r \to \infty$, 
which is the case if the spacetime is asymptotically flat, the frequency 
$\omega$ measured by a static observer becomes a function of $r$
by the gravitational redshift formula,
\begin{equation} \label{eq:gr-redshift}
\omega(r) = \frac{\omega_0}{\sqrt{A(r)}} \, .
\end{equation}
By (\ref{eq:H0}), a light ray with constant 
of motion $\omega _0$ is restricted to the region where
\begin{equation}\label{eq:restr}
\dfrac{\omega _0^2}{A(r)} \, > \,  \omega _p (r) ^2  \, .
\end{equation}
The restriction (\ref{eq:restr}) physically means that the photon 
frequency at a given point, $\omega(r)$, must be bigger than the plasma 
frequency, $\omega_p(r)$, at the same point. This is always true for
light propagation in a plasma.

To derive the orbit equation, we use (\ref{eq:Ham5}) and (\ref{eq:Ham6}) 
to find
\begin{equation}\label{eq:drdphi1}
\dfrac{dr}{d \varphi}  \, = \, 
\dfrac{\dot{r}}{\dot{\varphi}}
\, = \, 
\dfrac{D(r) p_r}{B(r) p_{\varphi}}   \, .
\end{equation}
Upon substituting for $p_r$ from (\ref{eq:H0}), this results in
\begin{equation}\label{eq:drdphi2}
\dfrac{dr}{d \varphi}  \, = \, 
\pm \, \dfrac{\sqrt{D(r)} }{\sqrt{B(r)}}
\sqrt{\dfrac{\omega _0^2}{p_{\varphi} ^2} \, h(r)^2 
\, - \, 1 \,}
\end{equation}
where we have defined the function
\begin{equation} \label{eq:h-definition}
h(r)^2 = \dfrac{D(r)}{A(r)} \left( 1 - A(r)  
\dfrac{\omega _p (r)^2}{\omega _0^2} \right) \, . 
\end{equation}
In general, the orbit has to be decomposed into sections where
$r$ is increasing as a function of $\varphi$ and sections where
it is decreasing, and the sign in (\ref{eq:drdphi2}) has to be 
chosen appropriately.  For a light ray that comes in
from infinity, reaches a minimum at a radius $R$, and goes out
to infinity again, integration over the orbit gives the formula 
for the bending angle, $\delta$, 
\begin{equation}\label{eq:delta}
\pi + \delta   \, = \, 2 \, \int _R ^{\infty} 
\dfrac{\sqrt{B(r)} }{\sqrt{D(r)} }
\left( \dfrac{ \omega _0^2}{p_{\varphi}^2}
\, h(r)^2 \, - \, 1 \,  \right)^{-1/2} 
\, dr   \, .
\end{equation}
As $R$ corresponds to the turning point of the trajectory,
the condition $dr/d \varphi \big | _R = 0$ has to hold. This
equation relates $R$ to the constant of motion 
$p_{\varphi}/\omega _0$,
\begin{equation}\label{eq:Rip}
h(R)^2 \, = \, \dfrac{p_{\varphi}^2}{ \omega _0^2}      \, .
\end{equation}
Then the deflection angle can be rewritten as a function of only $R$ and $\omega_0$ 
(for a given plasma distribution) as
\begin{equation}\label{eq:defl}
\pi + \delta   \, = \, 2 \, \int _R ^{\infty} 
\dfrac{\sqrt{B(r)} }{\sqrt{D(r)} }
\left( \frac{h(r)^2}{h(R)^2} - 1  \right)^{-1/2} 
\, dr   \, .
\end{equation}

\vspace{0.4cm}

\noindent
\emph{Example $1$: Schwarzschild spacetime}

\noindent
For the Schwarzschild spacetime,
\begin{equation}\label{eq:Ss}
A(r) = B(r)^{-1} = 1- \dfrac{2M}{r} \, , \quad 
D(r) = r^2 \, ,
\end{equation}
the function $h(r)$ specifies to
\begin{equation}\label{eq:Ssh}
h(r)^2 = r^2 \left( \dfrac{r}{r-2M} - \dfrac{\omega _p (r) ^2}{\omega _0^2} \right) \, .
\end{equation}
Then the bending angle reads
\begin{gather}\label{eq:Ssbending}
\pi + \delta   \, = 
\\
\nonumber
2 \int _R ^{\infty} \left( 
\dfrac{
r^2 \left( \tfrac{r}{r-2M}-\tfrac{\omega _p (r) ^2}{\omega _0^2} \right)
}{
R^2 \left( \tfrac{R}{R-2M}-\tfrac{\omega _p (R) ^2}{\omega _0^2} \right)
}
 -  1  \right) ^{\! \! \! -1/2} \! \! \! \! \! \! 
 \dfrac{dr}{\sqrt{r} \sqrt{r-2M}}
\, .
\end{gather}
This formula for the bending angle in a plasma on Schwarzschild spacetime was derived in \cite{Perlick2000b}. In \cite{TsBK2013} it was rederived using Synge's approach and 
rewritten in terms of an elliptic integral for a homogeneous plasma; there also the strong 
deflection limit ($\delta \gg 1$) was investigated.

\vspace{0.4cm}

\noindent
\emph{Example $2$: Ellis wormhole}

\noindent
As a second example we choose 
the Ellis wormhole \cite{Ellis1973b} which is a traversible
wormhole of the Morris-Thorne class \cite{MorrisThorne1988}.
It is true that the existence of such wormholes is questionable 
because they need exotic matter \cite{MorrisThorne1988} and,
at least for a certain kind of perturbations, it has been shown 
that the Ellis wormhole is unstable 
\cite{ShinkaiHayward2002,GonzalezGuzmanSarbach2009a,GoinzalezGuzmanSarbach2009b}.
On the other hand, wormholes have met 
with great interest because they make time travel possible
and the Ellis wormhole is an instructive example for
illustrating the applicability of our results. In this 
case the metric coefficients are
\begin{equation}\label{eq:Ellis}
A(r) = B(r) = 1 \, , \quad 
D(r) = r^2 + a^2 \, ,
\end{equation}
where the coordinate $r$ ranges from $- \infty$ to $\infty$ and 
$a$ is a constant that determines the radius of the ``neck'' of 
the wormhole. Note that this is an example of a spherically 
symmetric and static spacetime where we cannot make a transformation 
of the radius coordinate, $r \to \tilde{r}$, such that $\tilde{D}
( \tilde{r} ) =\tilde{r}{}^2$. The reason is that the function
$D(r)$ has vanishing derivative at $r=0$ (i.e., at the ``neck''),
so the desired transformation fails to be a good coordinate 
transformation on any radius interval that contains the point $r=0$.
For the Ellis wormhole the function $h(r)$ reads
\begin{equation}\label{eq:Ellish}
h(r)^2 = (r^2 +a^2) \left( 1- \dfrac{\omega _p (r) ^2}{\omega _0^2} \right)
\,,
\end{equation}
and the bending angle is given by
\begin{gather}\label{eq:Ellisbending}
\pi + \delta   \, = \, 
\\
\nonumber
2 \int _R ^{\infty} \left( 
\dfrac{
(r^2+a^2) \big( \omega _0 ^2 -\omega _p (r) ^2 \big)
}{
(R^2+a^2) \big( \omega _0 ^2  -\omega _p (R) ^2 \big)
}
 - 1 \right) ^{\! \! \! -1/2} \! \! \! \! \! \!
 \dfrac{dr}{\sqrt{r^2+a^2}}
\, .
\end{gather}


\section{Circular light orbits}\label{sec:circular}

We now derive the condition for cicular light orbits which
will be crucial for determining the shadow.
Along a circular light orbit we must have $\dot{r}=0$ and 
$\ddot{r}=0$. The first condition, by (\ref{eq:Ham6}), implies
$p_r=0$; from (\ref{eq:H0}) we get the equation
\begin{equation}\label{eq:circ1}
0 \, = \, - \dfrac{\omega _0^2}{A(r)}
+ \dfrac{p_{\varphi}^2}{D(r)}+
\omega _p (r) ^2   \,  .
\end{equation}
On the other hand, (\ref{eq:Ham6}) implies
\begin{equation}
\dot{p}_r = \frac{d}{d \lambda} 
\big( B(r) \, \dot{r} \big) = 
\ddot{r} B(r) + \dot{r}{}^2 B'(r) \, .
\end{equation}
>From this equation we read that $\dot{r}=0$ together with $\ddot{r}=0$ 
leads to $\dot{p}_r=0$, and from (\ref{eq:Ham3}) we get the second equation 
for circular light orbits,
\begin{equation}\label{eq:circ2}
0 \, = \, - \, 
\dfrac{\omega _0^2A'(r)}{A(r)^2}
\, + \, \dfrac{p_{\varphi}^2D'(r)}{D(r)^2}
\, - \, \dfrac{d }{dr} \omega _p (r) ^2  \,  .
\end{equation}
We solve each of these two equations (\ref{eq:circ1}) and 
(\ref{eq:circ2}) for $p_{\varphi}^2$,
\begin{equation}\label{eq:circ3}
p_{\varphi}^2 \, = \, D(r) \Big(
\dfrac{\omega _0^2}{A(r)}
- \omega _p (r) ^2 \Big)  \,  ,
\end{equation}
\begin{equation}\label{eq:circ4}
p_{\varphi}^2 \, = \, 
\dfrac{D(r)^2}{D'(r)} \Big( \dfrac{\omega _0^2 A'(r)}{A(r)^2}
+ \dfrac{d }{dr} \omega _p (r) ^2\Big)   \,  .
\end{equation}
Subtracting these two equations from each other yields,
after some elementary re-arrangements,
the equation for the radius of a circular light
orbit in the form
\begin{equation}\label{eq:circ8}
0 \, = \, \dfrac{d}{dr} h(r)^2 
\end{equation}
with the function $h(r)^2$ from (\ref{eq:h-definition}).
Any solution $r=r_{\mathrm{ph}}$ of (\ref{eq:circ8}) 
determines the radius of a \emph{photon sphere}. If a 
light ray starts tangentially to such a sphere it will stay 
on a circular path with radius $r_{\mathrm{ph}}$ forever. If the 
spacetime is asymptotically flat, and if $\omega _p (r) \to 0$
for $r \to \infty$, the outermost photon sphere is always
unstable with respect to radial perturbations. This means that 
the circular photon orbits in this photon sphere can serve as
limit curves for light rays that approach them asymptotically.
The radius $r_{\mathrm{ph}}$ of the outermost photon 
sphere is the critical value of the minimal radius $R$ mentioned
above. If a light ray comes in from infinity and reaches 
a minimum radius $R$ bigger than $r_{\mathrm{ph}}$, it will
go out to infinity again. The case $R=r_{\mathrm{ph}}$ 
corresponds to a light ray that spirals asymptotically 
towards a circular photon orbit in the sphere of radius 
$r_{\mathrm{ph}}$. All other rays cross the photon
sphere and we exclude the case that they can come back.
The latter case can occur only if there is a second
photon sphere.

In the vacuum case, $\omega _p (r) =0$, the
condition for a circular light orbit in an unspecified 
spherically symmetric and static spacetime was first given by
Atkinson \cite{Atkinson1965}. It is easy to check that
our condition (\ref{eq:circ8}) reduces, indeed, to
Atkinson's if $\omega _p (r) =0$.

\vspace{1.4cm}

\noindent
\emph{Example $1$: Schwarzschild spacetime}

\noindent
For the Schwarzschild spacetime, where $h(r)$ is given by (\ref{eq:Ssh}), condition
(\ref{eq:circ8}) for a photon sphere reads
\begin{equation}\label{eq:circSchw}
0 = \dfrac{r \, (r-3M)}{(r-2M)^2}
- \, \dfrac{\omega _p (r) ^2 }{\omega _0 ^2}
- \, r \, \dfrac{\omega _p (r) \omega _p ' (r)}{\omega _0 ^2} \, .
\end{equation}
For the special case that $\omega _p$ depends on $r$ via a
power law, this condition was already derived by 
Rogers \cite{Rogers2015}.
If there is no plasma, $\omega _p (r) =0$, (\ref{eq:circSchw})
gives the well-known result 
$r_{\mathrm{ph}}=3M$.
For the sake of curiosity, notice that  (\ref{eq:circSchw}) 
is \emph{identically} satisfied if 
\begin{equation}\label{eq:id}
h(r) ^2 = r^2  
\left( \dfrac{r}{r-2M} - \dfrac{\omega _p (r) ^2}{\omega _0^2} \right)
= C \, ,
\end{equation}
\begin{equation}
\text {i.e.} \: \;
\omega _p (r)^2 \, = \, \omega _0^2
\left( \dfrac{r}{r-2M} - 
\dfrac{C}{r^2} \right)
\end{equation}
with a positive constant $C$. For this particular density profile
of the plasma, there is a circular photon orbit with 
constant of motion $\omega _0$ at \emph{any} 
radius $r$ for which
\begin{equation}\label{eq:idr}
\dfrac{r^3}{r-2M} >  C \, .
\end{equation}

\vspace{0.4cm}

\noindent
\emph{Example $2$: Ellis wormhol}e

\noindent
For the Ellis wormhole, $h(r)$ is given by (\ref{eq:Ellish}) and condition (\ref{eq:circ8})
for a photon sphere reads
\begin{equation}\label{eq:Ellisps}
0 = \, r \left( 1 \, - \, \dfrac{\omega _p (r)^2}{\omega _0 ^2} \right)
- (r^2+a^2) \dfrac{\omega _p (r) \omega _p ' (r)}{\omega _0^2}
\, .
\end{equation}
Without a plasma, $\omega _p (r) = 0$, there is a unique photon sphere 
at the neck of the wormhole, $r_{\mathrm{ph}} = 0$. The same is true
for a homogeneous plasma, $\omega _p (r) = \mathrm{constant}$. However, 
in an inhomogeneous plasma there may be arbitrarily many photon spheres.
For an observer at a large positive $r$ coordinate, the outermost photon
sphere is relevant for the formation of the shadow which will be
discussed in the next section.

\section{Radius of the shadow}\label{sec:shadow}

\begin{center}
\begin{figure}
    \psfrag{x}{$\,$ \hspace{-1cm} $\sqrt{B(r)} dr$} %
    \psfrag{y}{$\,$ \hspace{-1.2cm} $\sqrt{D(r)} d \varphi$} %
    \psfrag{O}{$\,$} %
    \psfrag{T}{$\alpha$} %
    \psfrag{r}{$r_{\mathrm{ph}}$} %
\centerline{\epsfig{figure=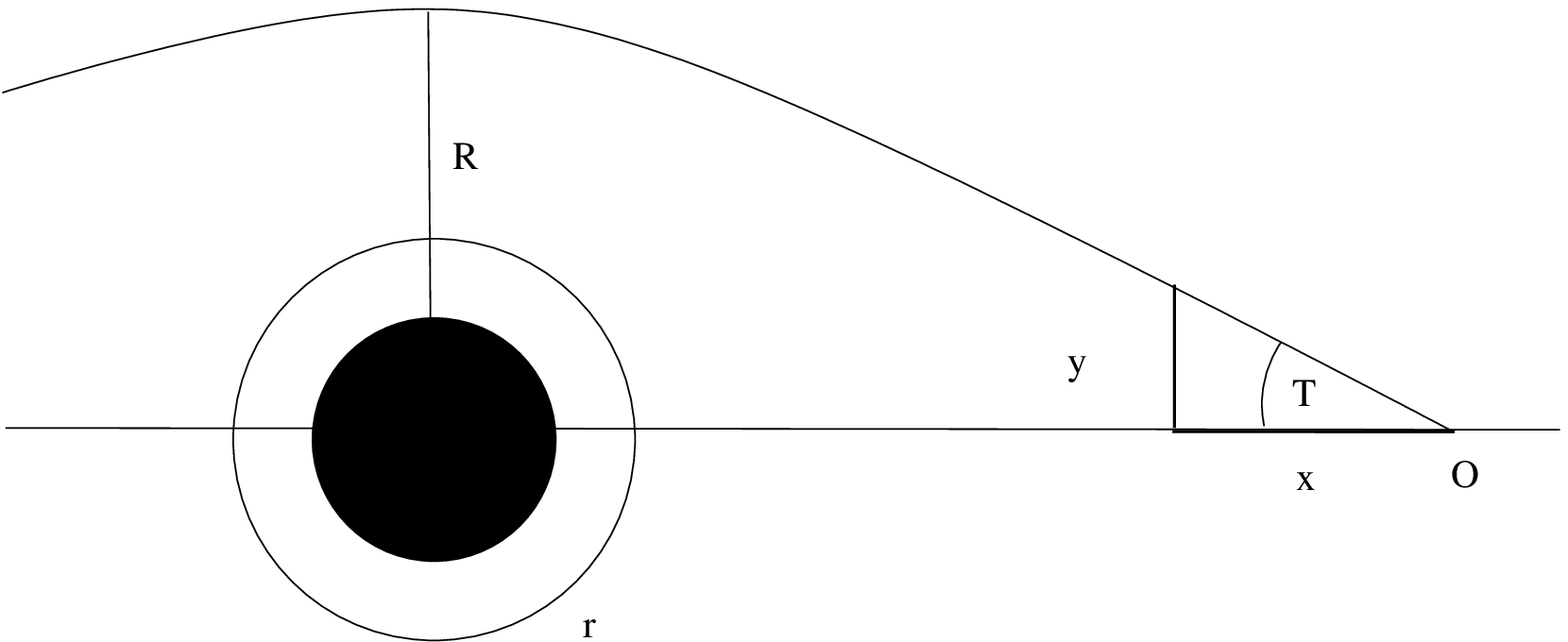, width=0.45\textwidth}}
\caption{For $R \to r_{\mathrm{ph}}$, the angle $\alpha$ approaches the 
angular radius $\alpha _{\mathrm{sh}}$ of the shadow.}
\label{fig:shadow}
\end{figure}
\end{center}

The shadow of a spherically symmetric and static black hole
is defined in the following way, see Fig.~\ref{fig:shadow}.
Consider light rays sent from an observer at radius coordinate 
$r_{\mathrm{O}}$ into the past. As we want to take the influence
of a plasma into account, by a light ray we mean a solution to
the equations of motion discussed in Section~\ref{sec:eom}.
These light rays can be divided into two classes: Light rays
of the first class go to infinity after being deflected by
the black hole. Light rays of the second class go towards 
the horizon of the black hole. If we assume that there are
no light sources between the observer and the black hole, 
initial directions of the second class correspond to darkness
on the observer's sky. This dark circular disk on the 
observer's sky is called the \emph{shadow} of the
black hole. The boundary of the shadow is determined
by the initial directions of light rays that asymptotically
spiral towards the outermost photon sphere. 
Here it is crucial that the light rays in the photon sphere 
are unstable with respect to radial perturbations because
otherwise they could not serve as limit curves. For this reason,
the construction of the shadow works for any spherically 
symmetric and static spacetime that admits an unstable  
photon sphere. This includes not only black holes
but also, e.g., wormholes. We have already mentioned that
in a spherically symmetric and static spacetime that is 
asymptotically flat the outermost photon sphere is 
always unstable, provided that the plasma density tends to
zero for $r \to \infty$. We will now calculate, for this 
situation, the angular radius $\alpha _{\mathrm{sh}}$ of 
the shadow.  The observer is assumed to be static somewhere 
between the outermost photon sphere and infinity.

We consider a light ray that is sent from the observer's
position at $r_{\mathrm{O}}$ into the past under an angle
$\alpha$ with respect to the radial direction. 
>From Fig.~\ref{fig:shadow} we read that $\alpha$ 
is given by
\begin{equation}\label{eq:alpha1}
\cot \, \alpha \, = 
\left. \frac{\sqrt{g_{rr}}}{\sqrt{g_{\varphi \varphi}}}  
\, \dfrac{dr}{d \varphi} \right|_{r=r_{\mathrm{O}}} = \left.
\dfrac{\sqrt{B(r)}}{\sqrt{D(r)}} \,
\dfrac{dr}{d \varphi} \right|_{r=r_{\mathrm{O}}} \, .
\end{equation}
If the light ray goes out again after reaching a minimum 
radius $R$, the orbit equation (\ref{eq:drdphi2}) can be 
rewritten, with the help of (\ref{eq:Rip}), as
\begin{equation}
\frac{dr}{d \varphi} = 
\pm \dfrac{\sqrt{D(r)}}{\sqrt{B(r)}} 
\sqrt{\dfrac{h^2(r)}{h^2(R)} \, - \, 1 \,} \, .
\end{equation}
For the angle $\alpha$ we obtain
\begin{equation}
\cot^2 \alpha \, = \frac{h^2(r_{\mathrm{O}})}{h^2(R)}  - 1 \, .
\end{equation}
Using
\begin{equation}
1 + \cot^2 \alpha = \frac{1}{\sin^2 \alpha}
\end{equation}
we get
\begin{equation}\label{eq:sinalpha}
\mathrm{sin} ^2 \alpha \, = \, \dfrac{h(R)^2}{h(r_{\mathrm{O}})^2}   \, .
\end{equation}
The boundary of the shadow $\alpha_{\mathrm{sh}}$ is determined by light 
rays that spiral asymptotically towards a circular light orbit at radius 
$r_{\mathrm{ph}}$. Therefore the angular radius of the shadow
is given by sending $R \rightarrow r_{\mathrm{ph}}$ in (\ref{eq:sinalpha}), 
\begin{equation}\label{eq:shadow}
\mathrm{sin} ^2 \alpha_{\mathrm{sh}} \, = \, 
\dfrac{h(r_{\mathrm{ph}})^2}{h(r_{\mathrm{O}})^2}  \, ,
\end{equation}
where $h(r)$ is given by the formula (\ref{eq:h-definition}).

For many applications we may assume that the observer is in a
region where the plasma density is negligibly small. Then
(\ref{eq:h-definition}) implies 
\begin{equation}\label{eq:hrO}
h(r_{\mathrm{O}})^2 = \dfrac{D(r_{\mathrm{O}})}{A(r_{\mathrm{O}})}
\end{equation}
and (\ref{eq:shadow}) reduces to
\begin{equation}\label{eq:shadowred}
\mathrm{sin} ^2 \alpha_{\mathrm{sh}} \, = \, 
\dfrac{D(r_{\mathrm{ph}}) A(r_{\mathrm{O}})}{A(r_{\mathrm{ph}}) D(r_{\mathrm{O}})}
\Big( 1- \dfrac{A(r_{\mathrm{ph}}) \omega _p(r_{\mathrm{ph}})^2}{\omega _0^2} \Big) 
\, ,
\end{equation}
\[
N(r_{\mathrm{O}}) \ll N(r_{\mathrm{ph}}) \, .
\]
This demonstrates that, under the assumptions stated, the 
plasma always has a \emph{decreasing} effect on the size of 
the shadow.

We emphasize that the preceding calculation applies not only to
black holes but also to other spherically symmetric and static
spacetimes with an unstable photon sphere, e.g. to ultracompact 
stars and to wormholes. As long as there is no light coming from 
the direction of the central object to the observer (which means, 
in particular, that the central object must not have a bright 
surface), these objects would cast a shadow in the same way as 
a black hole.

Let us summarize the results of this section. To find 
$\alpha_{\mathrm{sh}}$ for a given metric of the form (\ref{eq:g}), 
a given plasma concentration $N(r)$, a given photon frequency 
at infinity $\omega_0$ and a given observer position $r_{\mathrm{O}}$,
we have to calculate $r_{\mathrm{ph}}$ using eq.(\ref{eq:circ8}) and 
to substitute the result into formula (\ref{eq:shadow}).  Note that 
the photon frequency at the observer position is $\omega(r_{\mathrm{O}})$ 
according to (\ref{eq:gr-redshift}). 

\vspace{0.4cm}

\noindent
\emph{Example $1$: Schwarzschild spacetime}

\noindent
For the Schwarzschild case (\ref{eq:shadow}) specifies to
\begin{equation}\label{eq:Schwsh}
\mathrm{sin} ^2 \alpha _{\mathrm{sh}} =
\dfrac{
r_{\mathrm{ph}}^2  
\left( \dfrac{r_{\mathrm{ph}}}{r_{\mathrm{ph}}-2M} - 
\dfrac{\omega _p (r_{\mathrm{ph}}) ^2}{\omega _0^2} \right)
}{
r_{\mathrm{O}}^2  
\left( \dfrac{r_{\mathrm{O}}}{r_{\mathrm{O}}-2M} - 
\dfrac{\omega _p (r_{\mathrm{O}}) ^2}{\omega _0^2} \right)
}
\end{equation}
where $r_{\mathrm{ph}}$ has to be determined from 
(\ref{eq:circSchw}). 
In Section~{\ref{sec:accretion}} below we will 
evaluate this equation for a particular plasma density
profile. For vacuum, $\omega _p (r) = 0$, our consideration 
gives
\[
h(r)^2 = \frac{r^2}{1 - 2M/r} , \quad r_{\mathrm{ph}}=3M \, , 
\]
\begin{equation}\label{eq:Sssh}
\mathrm{sin} ^2 \alpha_{\mathrm{sh}} \, = \, 
\frac{27M^2 (1-2M/r_{\mathrm{O}})}{r_{\mathrm{O}}^2} \, .
\end{equation}
This is Synge's \cite{Synge1966} formula for the radius of the shadow 
of a Schwarzschild black hole which was mentioned already in the 
introduction. 

\vspace{1.4cm}

\noindent
\emph{Example $2$: Ellis wormhole}

\noindent
We now calculate the radius of the shadow for the Ellis wormhole.
In this case, our assumption of no light coming towards the observer
from the direction of the central object means, in particular,
that there are no light sources in the region $r < 0$. For the 
Ellis wormhole, the function $h(r)$ is given by (\ref{eq:Ellish}), 
so the equation (\ref{eq:shadow}) for the angular radius of the 
shadow specifies to
\begin{equation}\label{eq:Ellissh}
\mathrm{sin} ^2 \alpha_{\mathrm{sh}} \, = \, 
\dfrac{\big( r_{\mathrm{ph}}^2 +a^2 \big)}{\big( r_{\mathrm{O}}^2 +a^2 \big) }
\dfrac{
\big( \omega _0^2 - \omega _p (r_{\mathrm{ph}}) ^2 \big)
}{
\big( \omega _0^2 - \omega _p (r_{\mathrm{O}}) ^2 \big)
}
\, .
\end{equation}
For a homogeneous plasma, $\omega _p (r) = \mathrm{constant}$, we have 
$r_{\mathrm{ph}}=0$ and thus
\begin{equation}\label{eq:Ellisshhom}
\mathrm{sin} ^2 \alpha_{\mathrm{sh}} \, = \, 
\dfrac{a^2}{r_{\mathrm{O}}^2 +a^2}
\, ,
\end{equation}
so a homogeneous plasma has no influence on the size of the 
shadow.

\vspace{2cm}

\section{The shadow in a low-density plasma}\label{sec:thin}

If the plasma frequency is much smaller than the 
photon frequency, the equations for the photon 
sphere and for the radius of the shadow can be linearized about 
the corresponding values for vacuum light rays. To work this
out, we rewrite (\ref{eq:h-definition}) as
\begin{equation}\label{eq:epsilon}
h(r)^2=Q(r) \Big( 1 - \varepsilon \beta (r) \Big)
\end{equation}
where
\begin{equation}\label{eq:Qbeta}
Q(r) = \dfrac{D(r)}{A(r)} \, , \quad
\beta (r) = \dfrac{A(r) \omega_p (r)^2}{\omega _0^2} \, = \dfrac{\omega_p (r)^2}{\omega(r)^2} \, .
\end{equation}
Here we have introduced a book-keeping parameter $\varepsilon$
which will be set equal to unity after all equations have been
linearized with respect to $\varepsilon$.

The equation for a photon sphere, (\ref{eq:circ8}), now reads
\begin{equation}\label{eq:psweak}
0 = Q'(r) \Big( 1 - \varepsilon \beta (r) \Big)
- \varepsilon Q(r) \beta ' (r) \, .
\end{equation}
We write the solution to this equation in the form
\begin{equation}\label{eq:psepsilon}
r_{\mathrm{ph}} = r _{\mathrm{ph}} ^0 + \varepsilon r_{\mathrm{ph}} ^1 + \dots
\end{equation}
 where $r_{\mathrm{ph}} ^0$ is a solution in the case without plasma, i.e.,
\begin{equation}\label{eq:rph0}
Q'(r_{\mathrm{ph}} ^0) = 0 \, .
\end{equation}
After inserting these expressions into (\ref{eq:psweak}) and comparing
coefficients of $\varepsilon$ we find
\begin{equation}\label{eq:rph1}
r_{\mathrm{ph}}^1 =
\dfrac{Q(r_{\mathrm{ph}}^0) \beta '
(r_{\mathrm{ph}}^0)}{Q''(r_{\mathrm{ph}}^0)}
\, .
\end{equation}
Depending on the signs of $\beta '(r_{\mathrm{ph}})$ and $Q''(r_{\mathrm{ph}}^0)$,
this expression can be positive or negative, i.e., the plasma can shift the photon
sphere inwards or outwards. 

We now insert the expansion (\ref{eq:psepsilon}) into the equation for the shadow
(\ref{eq:shadow}). If we neglect all terms of quadratic and higher order in $\varepsilon$,
and set $\varepsilon$ equal to unity again, we find
\begin{equation}\label{eq:shepsilon}
\mathrm{sin} ^2 \alpha _{\mathrm{sh}} = \dfrac{Q( r_{\mathrm{ph}}^0)}{Q(r_{\mathrm{O}})}
\Big( 1- \beta ( r_{\mathrm{ph}}^0 ) + \beta (r_{\mathrm{O}}) \Big) \, , \quad \omega_p(r) \ll \omega(r) \, .
\end{equation}
Note that $r_{\mathrm{ph}}^1$ has dropped out from the equation for 
$\alpha _{\mathrm{sh}}$ to within this order. According
to (\ref{eq:shepsilon}), the plasma has a decreasing effect on 
the shadow as long as $\beta (r_{\mathrm{O}}) < \beta (r_{\mathrm{ph}}^0 )$.

Let us summarize the results of this section. To find 
the radius of the photon sphere $r_{\mathrm{ph}}$ for a
given metric (\ref{eq:g}), a given photon frequency 
at infinity $\omega_0$, and a given plasma frequency
that satisfies the condition $\omega_p(r) \ll \omega(r)$
one has to proceed in the following way: 
Write down the functions $Q(r)$ and $\beta(r)$, 
see (\ref{eq:Qbeta}); calculate $r_{\mathrm{ph}}^0$ 
from (\ref{eq:rph0}) and $r_{\mathrm{ph}}^1$ from 
(\ref{eq:rph1}); then calculate $r_{\mathrm{ph}}$ 
according to (\ref{eq:psepsilon}) with $\varepsilon=1$. 
To find the angular radius $\alpha _{\mathrm{sh}}$ of the 
shadow for a given observer position $r_{\mathrm{O}}$ 
in this case, we have to insert $r_{\mathrm{ph}}^0$ 
into (\ref{eq:shepsilon}).
 \vspace{1.9cm}

\noindent
\emph{Example $1$: Schwarzschild spacetime}

\noindent
As an example of the approximation formalism developed in
this section, we consider the Schwarzschild 
spacetime for the case that the plasma electron
density is given by a power law,
\begin{equation}
\dfrac{\omega _p (r) ^2}{\omega _0 ^2}  = \beta _0 \frac{M^k}{r^k} 
\end{equation}
where $\beta _0$ and $k$ are positive dimensionless constants. 
Then $r_{\mathrm{ph}}^0 = 3M$ and 
\begin{equation}
r_{\mathrm{ph}} ^1 =  \dfrac{\beta _0 \, M}{3^{k+1} } \, 
\left( 1- \frac{k}{2} \right) \, ,
\end{equation}
i.e., depending on the density profile the radius of the photon 
sphere may become smaller ($k>2$) or bigger ($k<2$) than $3M$. 
In the case that $k=2$ we obtain $r_{\mathrm{ph}} ^1=0$. Note 
that in this case the equation for the photon sphere can be 
solved exactly and gives the same result, $r_{\mathrm{ph}} = 3M$. 
The first-order equation for the radius of the shadow, 
(\ref{eq:shepsilon}), yields
\begin{gather}\label{eq:shadow-power}
\mathrm{sin} ^2 \alpha _{\mathrm{ph}} =
\\
\nonumber
\dfrac{27 M^2}{r_{\mathrm{O}}^2} \Big( 1 - \dfrac{2M}{r_{\mathrm{O}}} \Big) 
\left( 1- \dfrac{\beta _0}{3^{k+1}} 
+ \Big( 1 - \dfrac{2M}{r_{\mathrm{O}}} \Big) 
\dfrac{\beta _0 M^k}{r_{\mathrm{O}}^k}
\right) \, .
\end{gather}
If the observer is far away from the black hole, $r_{\mathrm{O}} \gg M$,
this can be simplified to
\begin{equation}\label{eq:shadow-distant}
\mathrm{sin} ^2 \alpha _{\mathrm{ph}} =
\dfrac{27 M^2}{r_{\mathrm{O}}^2} 
\left( 1- \dfrac{\beta _0}{3^{k+1}} \right) \, .
\end{equation}

\vspace{1.9cm}

\noindent
\emph{Example $2$: Ellis wormhole}

\noindent
For the Ellis wormhole, 
\begin{equation}\label{eq:Ellisld}
Q(r) = r^2+a^2 \, , \qquad
\beta (r) = \dfrac{\omega _p (r)^2}{\omega _0^2} \, .
\end{equation}
With $r_{\mathrm{ph}}^0=0$, the first-order approximation yields
\[
r_{\mathrm{ph}}^1= 
\dfrac{a^2 \omega _p(0) \omega _p ' (0)}{\omega _0^2}
\, ,
\]
\begin{equation}\label{eq:Ellisapprox}
\mathrm{sin}^2 \alpha _{\mathrm{sh}} =
\dfrac{a^2}{\big( r_0^2+a^2 \big)}
\left( 
1 - \dfrac{\omega _p(0)^2}{\omega _0^2}
+ \dfrac{\omega _p(r_{\mathrm{O}})^2}{\omega _0^2}
\right)
\, .
\end{equation}


\section{Spherically symmetric accretion of a plasma onto a Schwarzschild 
black hole}\label{sec:accretion}

In this section we consider in greater detail the special case
that the underlying spacetime is the Schwarzschild spacetime and 
that the plasma electron density corresponds to spherically 
symmetric accretion of dust. As usual, we use the word ``dust" 
as  a synonym for a pressure-less perfect fluid. Vanishing 
pressure implies that the flow lines are geodesics, i.e., that the
fluid particles are freely falling.
Note that for our plasma model the Hamiltonian and, 
thus, the  equations of motion for light rays do not depend on the
velocity of the plasma. This is, of course, different for other types 
of media, see Synge's book \cite{Synge1960}. In the case of an 
infalling plasma we have to determine the plasma density which is 
the only quantity that enters into the equation for the plasma 
frequency.

We start out from the continuity equation
\begin{equation}
\partial_i (\sqrt{-g} \rho u^i) = 0 \, .
\end{equation}
Here $\rho$ is the rest-mass density of the plasma, $u^i$ is the 4-velocity, and $g$ is 
the determinant of the metric. For the sake of simplicity, we consider a neutral hydrogen
plasma and we assume that the electrons have the same 4-velocity as the protons. As the 
electron mass is negligibly small in comparison to the proton mass, $\rho$ can be written 
as $\rho = m_p N$ where $m_p$ is the proton rest mass and $N$ is the number density
of the protons (which, as the plasma is assumed to be neutral, coincides with the number 
density of the electrons). 

For spherically symmetric and stationary accretion in
the Schwarzschild spacetime the continuity equation reduces to
\begin{equation}
\dfrac{d}{dr} \big( r^2 \rho (r) u^r (r) \big) = 0 \, .
\end{equation}
Integration gives
\begin{equation}\label{eq:mflux}
4 \pi  r^2 \rho (r)  u^r (r) = - \dot{M}_A = \mathrm{const}
\end{equation}
where $\dot{M}_A $ is a stationary mass flux.

We assume that the particles are dropped in radial free fall from rest at infinity, i.e., that 
our plasma is so ``cold'' that the pressure can be neglected. Then the integral curves of
$u^{\mu}$ are radial geodesics, so the radial component of the 4-velocity is 
\cite{HobsonEfstathiouLasenby2005}
\begin{equation}\label{eq:ursa0}
u^r (r) = \frac{dr}{d\tau} = - \sqrt{\frac{2M}{r}} \, .
\end{equation}
Thereupon, eq. (\ref{eq:mflux}) gives us the rest-mass density, 
\begin{equation}\label{eq:rhosa}
\rho (r) = \frac{\dot{M}_A}{4 \pi \sqrt{2M}} \frac{1}{r^{3/2}} \, .
\end{equation}
Equations (\ref{eq:ursa0}) and (\ref{eq:rhosa}) also follow from 
Michel's pioneering paper \cite{Michel1972} on fully relativistic 
spherically symmetric accretion where we have to specify
Michel's equations (9) and (10)  to the case of a dust. 

\begin{center}
\begin{figure}[h]
\centerline{\epsfig{figure=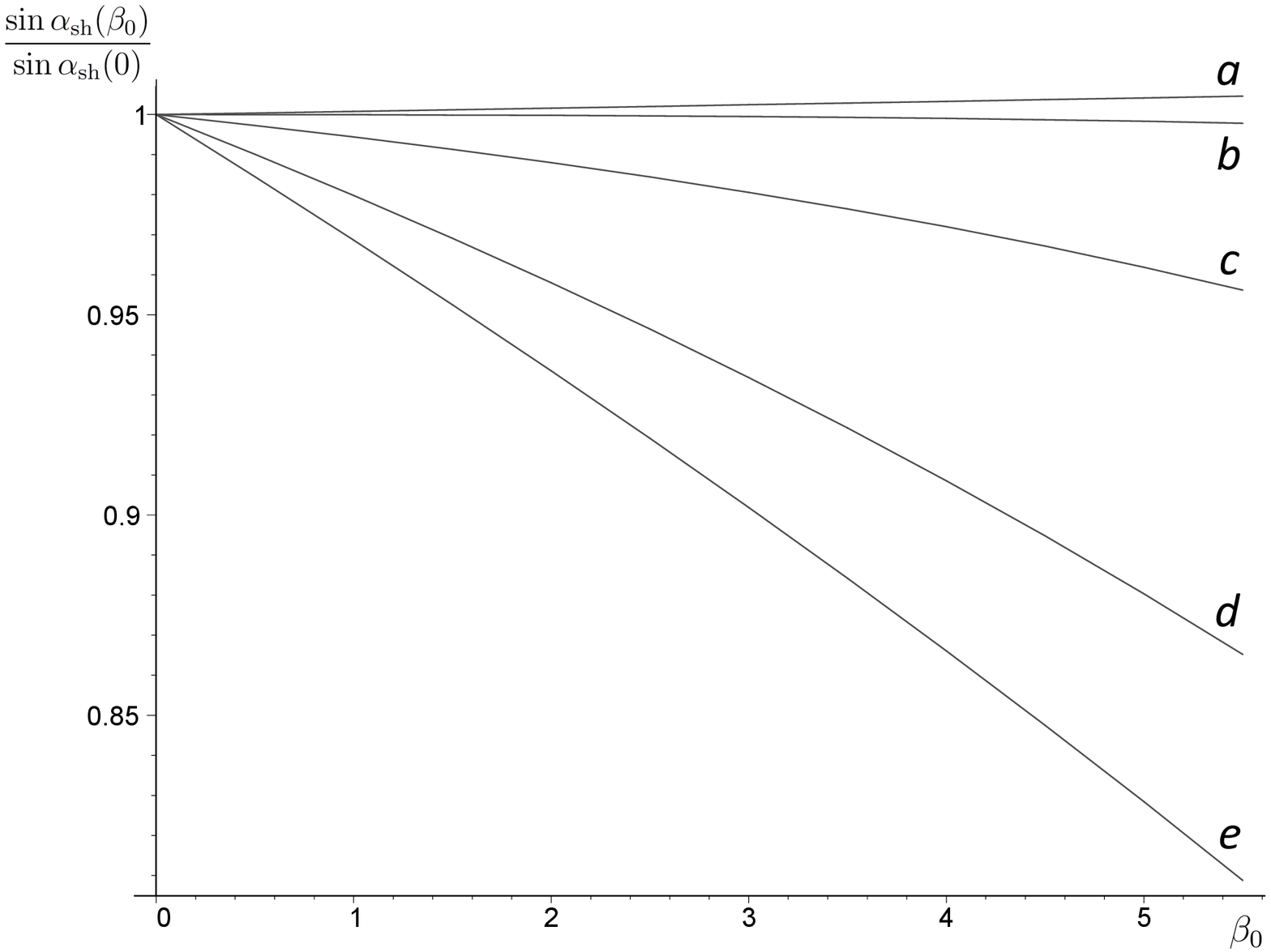, width=0.45\textwidth}}
\caption{Dependence of the radius of the shadow on $\beta_0$ for different 
observer positions $r_{\mathrm{O}}$: $r_{\mathrm{O}}/M = $ 3.333 (a), 
3.768 (b), 5 (c), 10 (d), 50 (e). The dependence is calculated using the
exact formulae. It is shown that for a distant observer the radius of
the shadow becomes smaller.}
\label{fig:exact}
\end{figure}
\end{center}

The rest-mass density $\rho (r)$ gives us the ratio of frequencies
\begin{equation}
\frac{\omega_p (r) ^2}{\omega_0^2} = \frac{4 \pi e^2 N(r)}{m_e \omega_0^2} 
=
\frac{4 \pi e^2 \rho(r)}{m_e m_p \omega_0^2} = \beta_0 \, 
\frac{M^{3/2}}{r^{3/2}} 
\end{equation}
where
\begin{equation} \label{eq:beta0}
\beta_0 = \frac{e^2 \dot{M}_A}{m_e m_p \omega_0^2  
\sqrt{2M} \, M^{3/2}} \, .
\end{equation}
For this function $\omega _p (r)$, the angular
radius (\ref{eq:Schwsh})  of the shadow is plotted, as a function of 
$\beta _0$ for different observer positions, in Fig.~\ref{fig:exact}.
 
Assuming that the approximation of a low-density plasma is justified, 
we can use formula (\ref{eq:shadow-power}) and find
\begin{gather}\label{eq:shaccr}
\sin^2 \alpha_{\mathrm{sh}} = 
\\
\nonumber
\frac{27M^2}{r_{\mathrm{O}}^2} 
\Big( 1- \dfrac{2M}{r_{\mathrm{O}}} \Big) 
\left( 1 - \frac{\beta _0}{3^{5/2}}  + 
\Big( 1- \dfrac{2M}{r_{\mathrm{O}}} \Big) 
\dfrac{\beta _0M^{3/2}}{r_{\mathrm{O}}^{3/2}} \right) \, .
\end{gather}
In this case the function $-\beta( r_{\mathrm{ph}}^0 ) + 
\beta(r_{\mathrm{O}})$ 
which enters into (\ref{eq:shepsilon}) is equal to zero when 
$r_{\mathrm{O}} \simeq 3.768 M$ and it takes its maximum when 
$r_{\mathrm{O}} = 3.333 M$. When $r_{\mathrm{O}} > 3.768 M$, 
the function is negative. Therefore, if the approximation of a 
low-density plasma is justified, the shadow becomes bigger if 
$r_{\mathrm{O}} < 3.768 M$ and it becomes smaller if 
$r_{\mathrm{O}} > 3.768 M$. 

As an aside, we mention that the 4-velocity component $u^r$
directly gives the velocity of the infalling particles as
measured by a static observer. To demonstrate this, 
following \cite{HobsonEfstathiouLasenby2005} (see also 
\cite{BK-Lovelace2002}), we first
observe that (\ref{eq:ursa0}) together with the normalization 
condition $g_{ij}u^iu^j=-1$ implies that  
\begin{equation}
u^0 = \frac{dt}{d\tau} = \left( 1 - \frac{2M}{r} \right)^{-1} \, ,
\end{equation}
hence
\begin{equation}
\frac{dr}{dt} = \frac{dr}{d\tau} \frac{d\tau}{dt} = 
- \sqrt{\frac{2M}{r}}  \left( 1 - \frac{2M}{r} \right) \, .
\end{equation}
For a static observer at radius coordinate $r$, a coordinate time 
interval $dt$ corresponds to a proper time interval
\begin{equation}
dt' = \left( 1 - \frac{2M}{r} \right)^{1/2}   dt \, .
\end{equation}
On the other hand, a radial coordinate separation $dr$ corresponds 
to a proper radial distance equal to
\begin{equation}
dr' = \left( 1 - \frac{2M}{r} \right)^{-1/2}  dr \, .
\end{equation}
As  a consequence, the velocity of the radially infalling particle, 
as measured by a static observer at $r$, is given by
\begin{equation}
\frac{dr'}{dt'} = \left( 1 - \frac{2M}{r} \right)^{-1}  \frac{dr}{dt}  
= - \sqrt{\frac{2M}{r}} = u^r \, .
\end{equation}\\
We will now use the results of this section for estimating
the numerial values of the plasma effects on the shadow of 
the supermassive black holes at the center of our galaxy 
and of M87.

For these applications we can always use the formulas with 
$r_\mathrm{O} \gg R_S$, so the angular radius of the shadow
(\ref{eq:shadow-distant}) can be written as
\begin{equation} \label{eq:main}
\sin^2 \alpha_{\mathrm{sh}} = \frac{27R_S^2}{4 r_{\mathrm{O}}^2} 
\left( 1 - Z \right) \, , \quad Z = \frac{\beta _0}{3^{5/2}} \, .
\end{equation}
$\beta_0$ was defined in (\ref{eq:beta0}). For estimating 
the plasma correction numerically, we re-express all quantities
in Gaussian cgs units, i.e., we restore factors of $c$ and $G$.
Then $\beta _0$ reads
\begin{equation}
\beta_0 = \frac{e^2 \dot{M}_A c^3}{m_e m_p \omega_0^2  \sqrt{2GM} (GM)^{3/2}} \, .
\end{equation}
The mass accretion rate $\dot{M}_A$ can be estimated via the 
observed luminosity of the galactic center. We write
\begin{equation}
L \simeq \eta \dot{M}_A c^2 \, ,
\end{equation}
where $\eta$ is a non-dimensional coefficient characterizing the 
accretion efficiency. We obtain for the correction due to the presence
of the plasma:
\begin{equation}
Z = 
\frac{2^{3/2}}{3^{5/2}} \frac{e^2}{m_e m_p c} \frac{L}{\eta c^2} \frac{1}{R_S^2 \omega_0^2} .
\end{equation}
Note that the circular frequency $\omega_0$ is connected with the 
ordinary frequency $\nu_0$ by the usual relation, $\omega_0 = 2 \pi \nu_0$,
and that in the expression for $Z$ we can assume, as a valid approximation,
that $\omega _0$ is related to the wavelength $\lambda _0$ by the
vacuum relation $\omega_0 = 2 \pi c / \lambda_0$.

The value of $\eta$ depends on the specific model of accretion. 
For estimations we consider $\eta_1 = 10^{-4}$ (see \cite{Narayan1994}, 
\cite{Narayan1995}) and $\eta_2 = 0.1$ (see \cite{BKLovelace1997}).

Let us consider the galactic center source Sgr A*. For estimations we put 
$M = 4.3 \times 10^6 M_{\odot}$, $r_\mathrm{O} = 8.3$ kpc 
(see \cite{Ghez2008}, \cite{Gillessen2009}). Then Synge's formula
(\ref{eq:Sssh}) gives for the angular radius of the shadow in vacuum
a value of $\alpha_\mathrm{sh} \simeq 27$ $\mu$as, i.e., a 
diameter of about 53 $\mu$as which is expected to be resolvable with 
Very Long Baseline Interferometry (VLBI) soon \cite{Doeleman2008}, 
\cite{Huang2007}. By contrast, a naive Euclidean estimate of the 
horizon size, $\alpha \approx \mathrm{tan} \, \alpha = R_S/r_{\mathrm{O}}$,
yields $\alpha \approx 10 \, \mu$as, i.e., a diameter of 
about 20 $\mu$as, as was already mentioned in the introduction. For a rotating (Kerr) black hole, the shadow is 
flattened on one side, but its vertical diameter is still given 
by Synge's formula if $r_{\mathrm{O}} \gg R_S$, see
\cite{GrenzebachPerlickLaemmerzahl1915}.

For estimating the correction $Z$ we use $L = 10^6 L_{\odot}$ 
(see, for example \cite{Davidson1992}). For $\lambda_0 = 1$ mm 
we obtain:
\begin{equation}
Z = 0.8 \times 10^{-5} \; \mbox{for} \; \eta_1, \quad  
Z = 0.8 \times 10^{-8} \;\mbox{for} \; \eta_2 .
\end{equation}
Note that observations of the shadow are planned in the (sub)millimeter 
regime because at wavelengths of more than about 1.3 mm it is 
expected that the shadow is washed out by scattering
\cite{FalckeMeliaAgol2000}. We see that in this
regime the effect of a plasma is rather small, for the chosen
values of $\eta$. However, for bigger 
radio wavelengths the effect can be significant. For $\lambda_0 = 10$ 
cm we obtain:
\begin{equation}
Z = 0.8 \times 10^{-1} \; \mbox{for} \; 
\eta_1, \quad Z = 0.8 \times 10^{-4} \;
\mbox{for} \; \eta_2 .
\end{equation}

We now turn to the galactic center source M87. For estimations we 
put $M = 3 \times 10^9 M_{\odot}$, $r_\mathrm{O} = 18$ Mpc, 
$L = 7 \times 10^{40}$ ergs s$^{-1}$ (see, for example 
\cite{Davidson1992}). Then Synge's formula leads to an 
angular radius of the shadow in vacuum of $\alpha_\mathrm{sh} 
\simeq 9$ $\mu$as.

For $\lambda_0 = 1$ mm we obtain:
\begin{equation}
Z = 0.3 \times 10^{-9} \; \mbox{for} \; \eta_1, 
\quad  Z = 0.3 \times 10^{-12} \;\mbox{for} \; \eta_2 .
\end{equation}
For $\lambda_0 = 100$ cm we obtain:
\begin{equation}
Z = 0.3 \times 10^{-3} \; \mbox{for} \; 
\eta_1, \quad  Z = 0.3 \times 10^{-6} \;\mbox{for} \; \eta_2 .
\end{equation}


\section{Conclusions}

In this paper the first steps towards an investigation of the shadow in the
presence of matter based on analytical calculations have been taken. We have 
analytically calculated the influence of a non-magnetized pressure-less 
plasma on the angular size of the shadow. It was our goal to derive
all relevant formulas for an unspecified spherically symmetric and 
static spacetime, before specializing to the case of a Schwarzschild 
black hole. The gravitational field was not supposed to be weak. We 
have worked in the frame-work of geometrical optics; in this approximation 
the presence of the plasma leads to a change of the geometrical size of 
the shadow via a change of the light rays in this medium.

The equation of motion for light rays in a plasma with a spherically 
symmetric density distribution on a spherically symmetric and static 
spacetime (\ref{eq:g}) has been derived, see (\ref{eq:drdphi2}). 
In particular, the formula for the photon deflection angle has been shown 
for unbound orbits, see (\ref{eq:defl}). We have also found the radii 
of the circular light orbits (radii of the photon spheres), see 
(\ref{eq:circ8}). The central result of our paper is an analytical 
formula for the angular size of the shadow, 
see (\ref{eq:shadow}).
\footnote{Soon after a preprint of the present paper
was made public on the arXiv \cite{our2015}, a paper by
Atamurotov and Ahmedov \cite{Ahmedov2015} appeared, again
on the arXiv, where the shadow was calculated for the case
of a \emph{homogeneous} plasma. Their results are in obvious
disagreement with ours if the latter are specialized to the
case of a homogeneous plasma. The reason is in the fact that
they erroneously treated the index of refraction as a
constant. Actually, the index of refraction involves the
plasma density, which is constant in a homogeneous plasma,
and the photon frequency, which in a gravitational field
depends on the space point. Therefore, the index of
refraction is \emph{not} a constant for a homogeneous
plasma. This has the consequence that in the Schwarzschild
spacetime the photon sphere in a homogeneous plasma is
different from the vacuum case. In the meantime, a strongly revised version of
their paper appeared in PRD \cite{Ahmedov2015a} where they are
now considering a plasma density proportional
to $r^{-1}$.} 

Special attention was given to the realistic case when the plasma 
frequency is much smaller than the photon frequency, see 
Section \ref{sec:thin}. We have shown that in this case
the plasma has a decreasing effect on the size of the shadow
provided that the plasma density is higher at the photon sphere 
than at the observer position, see (\ref{eq:shepsilon}). In particular, 
a simple formula for the size of the shadow was presented for 
the case of a power-law density distribution and an observer
at a large distance, see (\ref{eq:shadow-distant}).

In the presence of a plasma the size of the shadow depends on the 
wavelength at which the observation is made, in contrast to the vacuum 
case where it is the same for all wavelengths.
For the underlying spacetime we have treated two examples, the 
Schwarzschild black hole and the Ellis wormhole. In particular,
the case of spherically symmetric accretion of plasma onto a 
Schwarzschild black hole was considered in detail, see 
Section \ref{sec:accretion}. We have found that for an
observer far away from the Schwarzschild black hole the plasma
makes the shadow smaller. As examples, we have considered Sgr A* and M87. 
For the specific accretion model used here we have found that
the effect of the presence of a plasma on the size of the shadow 
can be significant only for wavelengths of at least a few centimeters. 
At such wavelengths the observation of the shadow is made difficult 
because of scattering. 

Apart from considering more complicated
plasma models, an obvious next step would be 
to generalize the analysis to the case of 
axially symmetric and stationary situations
to include rotating black holes. Then the 
shadow is no longer circular, and the plasma has an effect not
only on its size but also on its shape. We are planning to work
out the details in future work.

\section*{Acknowledgments}
VP wishes to thank Deutsche Forschungsgemeinschaft for financial 
support under Grant No. LA 905/14-1 and the Dynasty Foundation 
for financing his visit to Moscow where part of this work was carried 
through. Also, VP acknowledges support from the Deutsche 
Forschungsgemeinschaft within the Research Training Group 1620 
``Models of Gravity''. The results for the 
Schwarzschild black hole were obtained by OYuT and GSBK, and 
this part of work was financially supported by Russian 
Science Foundation, Grant No. 15-12-30016.

\bibliographystyle{ieeetr}

\end{document}